\documentclass[preprint,authoryear,12pt]{elsarticle}

\usepackage{graphicx}
\usepackage{wrapfig}

\usepackage{amssymb}

\journal{High Energy Density Physics Journal (HEDP)}

\begin{document}

\begin{frontmatter}



\title{Comparing Poynting flux dominated magnetic tower jets 
with kinetic-energy dominated jets}

\author[1]{M. Huarte-Espinosa}
\author[1]{A.~Frank}
\author[1]{E.~G.~Blackman}
\author[2,3]{A.~Ciardi}
\author[4]{P.~Hartigan} 
\author[5]{S.~V.~Lebedev}
\author[5]{J.~P.~Chittenden}
\address[1]{Department of Physics and Astronomy, University
of Rochester, 600 Wilson Boulevard, Rochester, NY, 14627-0171}
\address[2]{LERMA, Universit\'e Pierre et Marie Curie,
Observatoire de Paris, Meudon, France}
\address[3]{\'Ecole Normale Sup\'erieure, Paris, France. UMR 8112 CNRS}
\address[4]{Rice University, Department of Physics and Astronomy, 6100 S. Main,
Houston, TX 77521-1892}
\address[5]{The Blackett Laboratory, Imperial College London, SW7 2BW London, UK}

\begin{abstract}
Magnetic Towers represent one of two fundamental forms of MHD
outflows. Driven by magnetic pressure gradients, these flows have
been less well studied than magneto-centrifugally launched jets even
though magnetic towers may well be as common. Here we 
present new results exploring the behavior and evolution of magnetic
tower outflows and demonstrate their connection with pulsed power
experimental studies and purely hydrodynamic jets which
might represent the asymptotic propagation regimes of 
magneto-centrifugally launched jets.
High-resolution AMR MHD simulations (using
the AstroBEAR code) provide insights into the underlying physics of
magnetic towers and help us constrain models of their propagation.
Our simulations have been designed to explore the effects of thermal
energy losses and rotation on both tower flows and their hydro counterparts. 
We find these parameters have significant
effects on the stability of magnetic towers, but mild
effects on the stability of hydro jets. Current-driven 
perturbations in the Poynting Flux Dominated (PDF) towers 
are shown to be amplified in both the cooling and rotating cases. 
Our studies of the long term evolution of the towers show that 
the formation of weakly magnetized central jets within the tower are broken up by
these instabilities becoming a series of collimated clumps which
magnetization properties vary over time. 
In addition to discussing these
results in light of laboratory experiments, we address their
relevance to astrophysical observations of young 
star jets and outflow from highly evolved solar type stars.
\end{abstract}


\end{frontmatter}



\section{Introduction} 

Astronomical observations have established that collimated supersonic
outflow, or jets, are ubiquitous. Young Stellar Objects (YSO),
post-AGB stars, X-ray binaries and active radio galaxies show jets.
The ``central engines'' of these flow cannot be directly observed
due to insufficient telescope resolution.  The launch and collimation
of jets has been modelled in terms of a combination of accretion,
rotation and magnetic mechanisms (Pudritz et al.~\citeyear{pudritz07}).
The study of magnetized supersonic jets has recently reached the
laboratory, where experiments have provided scale models
of the launch and propagation of magnetized jets with dimensionless
parameters relevant for astrophysical systems (Lebedev
et~al.~\citeyear{lab1}; Ciardi et~al.~\citeyear{ciardi9}; Suzuki-Vidal
et~al.~\citeyear{suzuki}).  In combination with numerical simulations
(Ciardi et~al.~\citeyear{ciardi7}; Huarte-Espinosa et~al.~\citeyear{we}),
these experiments help us to resolve unanswered questions on jet
physics. In particular the distinction between the physics of jet
launch and that of jet propagation far from the engine. Since we
cannot observe the former, it is important to identify distinct
features of jets in the asymptotic propagation regime that can
distinguish different engine paradigms. While both simulations and
experiments now consistently reveal the promise, if not essentiality,
of dynamically significant magnetic fields for jet launch, the
correlation between the initial jet magnetic configuration and the
stability of the flow far from the launching region is unclear.

The importance of the magnetic field flux relative to the
flows' kinetic energy divides jets into (i) Poynting flux dominated
(PFD; \citealp{shibata86}; \citealp{bell96}; \citealp{ustyugova00};
\citealp{lovelace02}; Nakamura \& Meier, 2004), in which magnetic
fields dominate the jet structure, (ii) magneto-centrifugally launched
(MCL; \citealp{blandford82}; Ouyed \& Pudritz, 1997; \citealp{blackman01};
\citealp{mohamed07}), in which magnetic fields only dominate out
to the Alfv\'en radius.  The observable differences between PFD and
MCL jets are unclear, as are the effects that cooling
and rotation have on PFD jets.

\section{Laboratory experiments}
\label{exps}
\vskip-.3cm

Lebedev et~al.~(\citeyear{lab1}), Ciardi et al. (2007,2009) and
Suzuki-Vidal et~al. (2010) have carried out laboratory experiments
of magnetized jets in the pulsed power facilities of Imperial College
London. Lebedev et~al.~(\citeyear{lab1}) adjusted the dimensionless
numbers (Reynolds, magnetic Reynolds and Peclet) of the plasma
in these experiments in appropriate regimes for astrophysics. 
A critical ingredient of these laboratory
experiments is the significant
thermal energy loss of both the jets and the ambient medium plasmas;
it plays a critical role in many astrophysical jet environments, e.g. YSOs.
The
experimental setup of Lebedev et~al.~(\citeyear{lab1}) consisted of a pair of
concentric electrodes connected by a conical array of tungsten
wires, of 13\,$\mu$m in diameter, inside a vacuum chamber. A TW
electrical pulse (1\,MA, 250\,ns) was applied to this circuit. Such
current causes ablation of the wires which results in the formation
of a background ambient plasma (Figure~1a).  This material is pushed
above the wires by Lorentz forces then, while resistive diffusion
keeps the current close to the wires. The current induces a toroidal
magnetic field which at this stage is confined around the central
electrode, below the wires. 
After the complete ablation of the
section of the wires near the central electrode, the current switches
to the plasma and creates a magnetic cavity (Figure~1b) containing a
central jet.  The jet's core is confined and accelerated upward by
toroidal magnetic field pressure. The return current flows along
the walls of the magnetic cavity which is in turn confined by both
the thermal pressure and the inertia of the ambient medium plasma.

\begin{figure}[hb]
\vspace{-00pt}
\begin{center}
\includegraphics[width=\textwidth]{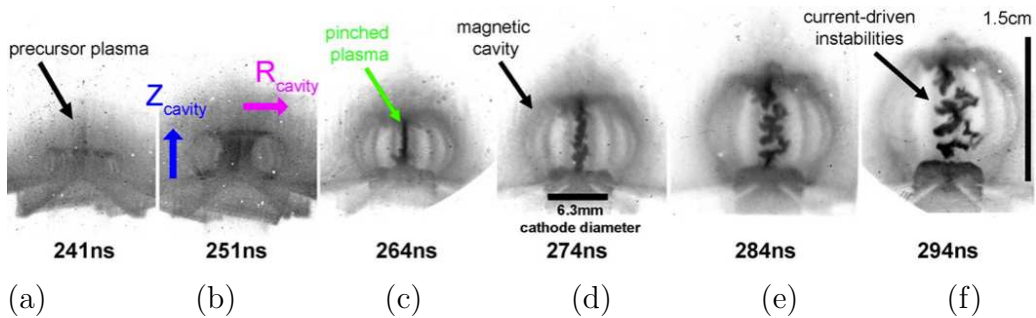}
~~~~~~~~
(a)
~~~~~~~~~~~~~~
(b)
~~~~~~~~~~~~~~
(c)
~~~~~~~~~~~~~~
(d)
~~~~~~~~~~~~~~
(e)
~~~~~~~~~~~~~~
(f)
~~~~~~ \\
\vspace{-0pt}
\caption{Soft X-rays images showing a time sequence of the formation
of the background plasma (a), the expansion of the magnetic cavity (b), the
launch of the jet (c), the development of instabilities in the central
jet column (d) and the fragmentation of the jet (e,f). 
The wire array shows in the bottom part of the figures.
Image taken from Suzuki-Vidal et~al.~(\citeyear{suzuki}).}
\end{center}
\vspace{-08pt}
\end{figure}

Then, the magnetic cavity opens up, the jet becomes detached and
it propagates away from the source, in a collimated fashion, at
velocities of order 200\,km\,s$^{-1}$ (Figure~1c). 

Lebedev et~al.~(\citeyear{lab1}) found that the body of the jets in
their experiments was significantly affected by the development of
instabilities (Figure~1e). The outflows where fragmented into a well
collimated set of clumps, or ``bullets'', with characteristic axial
non-uniformities (Figure~1e,f; Lebedev et~al.~\citeyear{lab1}).  Ciardi
et~al.~(\citeyear{ciardi7}) reproduced these experiments using 3D
non-ideal MHD numerical simulations carefully designed to model the
laboratory components (electrodes and wires) and all the plasma
evolution phases. The simulations of Ciardi et~al.~(\citeyear{ciardi7})
reproduced the experimental results of Lebedev et~al.~(\citeyear{lab1})
very well. The simulations clearly showed that during the final
unstable phase of jet propagation the magnetic fields in the central
jet adopted a twisted helical structure. Thus Ciardi
et~al.~(\citeyear{ciardi7}) concluded the nano-metric jets of Lebedev
et~al.~(\citeyear{lab1}) are affected by normal $m=\,$1 mode perturbations.

\section{Simulations}
\label{model}

We use the Adaptive Mesh Refinement code AstroBEAR2.0 (Cunningham
et~al.~\citeyear{bear}; Carroll-Nellenback et~al.~\citeyear{bear2}) to solve
the equations of MHD in 3D with cooling source terms.  The grid represents
160$\times$160$\times$400\,AU divided into 64$\times$64$\times$80
cells plus 2 adaptive refinement levels.  We use extrapolation boundary
conditions at the four vertical faces of the domain and at the top
one, as well as a combination of MHD reflective and
inflow conditions at the bottom face.

We will now briefly describe the setup of our simulations,
see Huarte-Espinosa et~al.~(2012b)
for a full description of the implementation.
Initially, the gas is static and has an ideal gas equation of state
($\gamma=\,$5$/$3), a number density of 100\,cm\,s$^{-1}$ and a
temperature of 10000\,K.  The magnetic field is helical, centrally
localized (${\bf B}\ne 0$ within $r,z<r_e$) and the magnetic pressure exceeds the thermal pressure
inside the magnetized region. 
%
We use source terms to continually inject magnetic or kinetic energy
at cells where $r,z<r_e$. We have carried out six simulations: three PFD
magnetic tower jets and three hydrodynamic jets. For each case we
calculated an adiabatic, a cooling and a rotating case. The
hydrodynamic runs were implemented is such as way as to have same
time average propagation speed and energy flux as the adiabatic PFD
jet run. We use the cooling tables of Dalgarno \& McCray~(\citeyear{dm})
for the cooling case. For the rotating runs we applied a Keplerian
rotation profile to the gas and magnetic fields located within
$r,z<r_e$. 

\section{Results}

\subsection{Jet structure and stability}

Our magnetic tower simulations consistently show that magnetic
pressure gradients, alone, push field lines and plasma upward,
forming magnetic cavities with low densities (Figure~2, all but
right panel). PFD jets decelerate relative to the hydro ones;
although both the PFD and the hydro jets have the same injected
energy flux, the PFD case produces both axial and radial expansion
due to magnetic pressure.  \\

\begin{wrapfigure}[22]{r}{.73\textwidth}
\vspace{-30pt}
\begin{center}
\includegraphics[width=.70\textwidth]{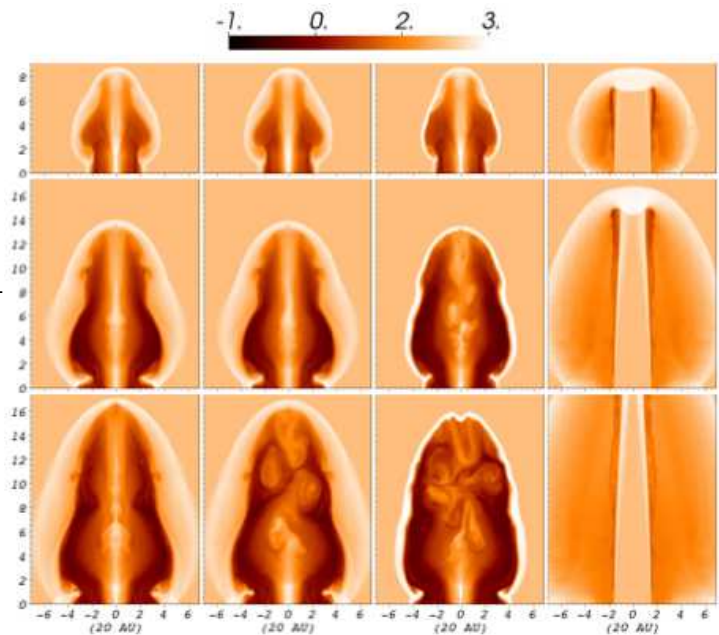} \\
\vspace{-10pt}
\caption{Logarithmic false color density maps of the adiabatic 
(1st column), rotating
(2nd column) and cooling (3rd column) PFD jets. Hydrodynamic jet 
(4th column). From top to bottom the time is 42, 84 and 118\,yr.} 
\end{center} 
\vspace{+00pt}
\end{wrapfigure}

The cores of the PDF jets are confined
by magnetic hoop stress, while their surrounding cavities are
collimated by external thermal pressure.  In contrast, the
pre-collimated hydro jets can only expand via a much lower thermal
pressure. Thus for our setup all of the energy flux in the
hydro-case is more efficiently directed to axial mechanical power.
It is worth mentioning that our hydro cases can emulate to some
degree the asymptotic propagation regime of a jet that was
magneto-centrifugally launched (e.g. Blackman 2007) which is distinct
from a PFD jet. \\

Regarding the structure of the jets, we find that the PFD jet cores are
thin and unstable, whereas the hydro jet beams are thicker, smoother
and stable. These are observationally distinguishable features.
The base rotation and the cooling conditions that we explore (see end
of Section~\ref{model}) have mild effects on the propagation of the
hydro jets, but strong effects on the stability of PFD jets. \\

Instabilities in our PDF jets are current driven; the jets carry
high axial currents which return along their outer contact
discontinuity.  Lebedev et~al.~(\citeyear{lab1}) saw this current density 
distribution in their magnetic tower experiments.  
We find that the thermal to magnetic pressure ratio
distribution, $\beta({\bf X},t)$, consists of a central high beta
plasma column --the jets' core-- surrounded by the low beta plasma of the
cavities (Figure~3). 
The growth rate of the current driven
instabilities is accelerated by cooling, firstly, and by base
rotation, secondly.  

\begin{figure}[hb]
\begin{center}
$\log (\beta)$ \\
\includegraphics[width=.9\columnwidth]{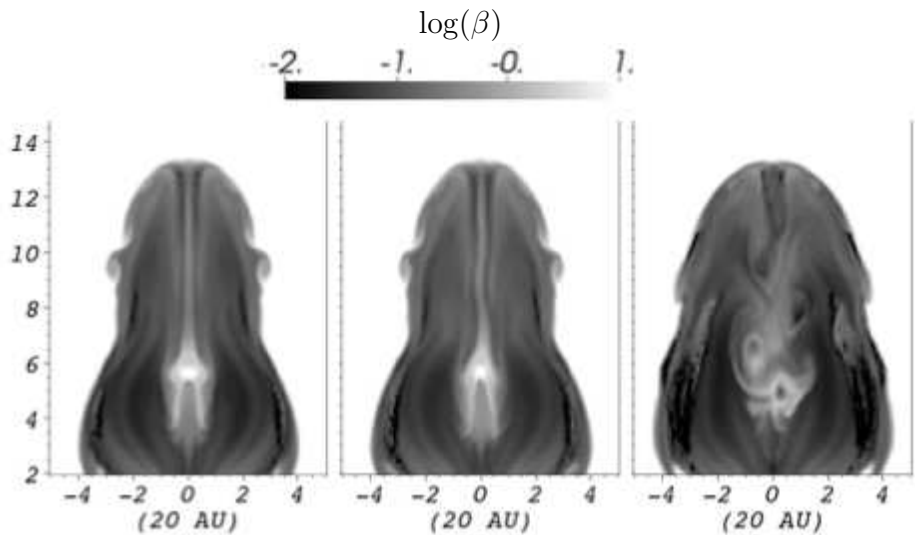} \\
\caption{Thermal to magnetic pressure ratio of out PDF magnetic
towers at time$=$84\,yr.  These are logarithmic false gray-scale
maps of the adiabatic (left), the rotating (middle) and the cooling
(right) jets.}
\label{beta}
\end{center}
\end{figure}

At the core of these jets the axial
magnetic field dominates over the toroidal one. $|B_{\phi} / B_z |
\ll 1$.  Thus the columns' instability condition is given by
\begin{equation}
   \left| \frac{B_{\phi}}{B_z} \right| > | (\beta_z - 1)k r_{jet}  |,
   \label{insta}
\end{equation}
\noindent where $\beta_z=2 \mu_0 P / B_z^2$, $\mu_0$ is the magnetic
vacuum permeability, $P$ is the plasma thermal pressure and $k^{-1}$
is the characteristic wavelength of the current-driven perturbations
(Huarte-Espinosa et~al., 2012b).

The cooling jet core consistently shows $\beta_z \sim\,$1
which means that because of thermal energy losses it does not have sufficient
thermal energy to damp the toroidal magnetic pressure. Hence kink perturbations
grow exponentially.
A different path to instability operates in the case in which we
apply a Keplerian rotation profile at the base of the PDF jet. Rotation
causes a slow amplification of the toroidal magnetic field, hence
the left hand side of equation~(\ref{insta}) increases slowly and
so do the kink mode perturbations. The rotating jet is not completely
destroyed by these perturbations, and their amplitude is about twice
the radius of the central jet (see Huarte-Espinosa et~al., 2012b
for details), in agreement with the Kruskal-Shafranov criterion
(Kruskal et~al.~\citeyear{kruskal}; Shafranov~\citeyear{shafranov}).

\subsection{Energy flux distribution}

\begin{wrapfigure}[38]{r}{.5\textwidth}
\vspace{-15pt}
\begin{center}
~~~~~~~~~~~$\log \left| Q({\bf x},t) \right|$ \\
\includegraphics[width=.4\columnwidth]{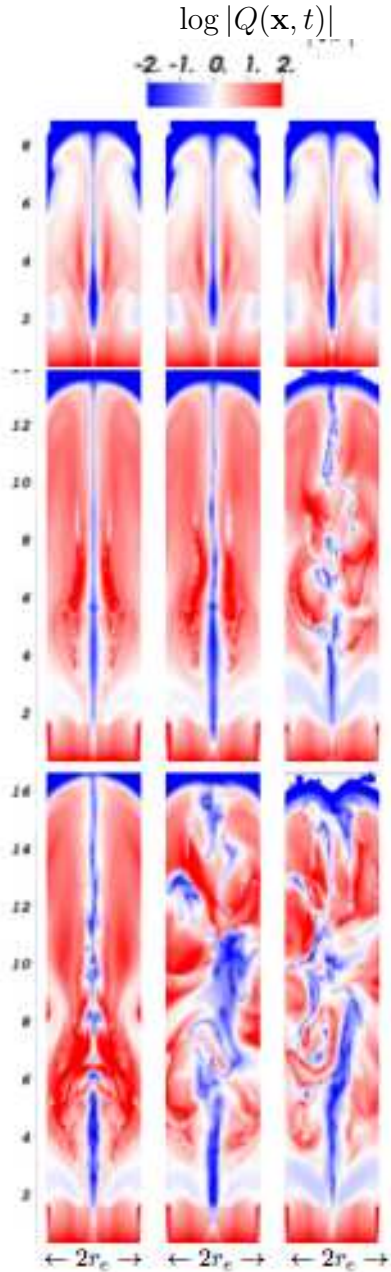} \\
\caption{Logarithmic color maps of the distribution and evolution
of the magnetic towers Poynting to kinetic flux ratio.}
\label{pfd}
\end{center}
\end{wrapfigure}

In Figure~\ref{pfd} we show logarithmic color maps of the distribution
of the jet Poynting to kinetic flux ratio $Q({\bf x},t)= f_P/f_k$
as a function of time, where 
\begin{equation}
   \begin{array}{c l l}
   f_P = & \int\limits_s \, [{\bf B \times (V \times B ) }]_z \, dS, \\
   f_k = & \int\limits_s  \, \frac{1}{2} \, \rho \, |{\bf V}|^2 \, V_z \, dS.
   \end{array}
\label{fluxratio} 
\end{equation}
\noindent These integrals are taken over the area of the jets' beam.

We consistently see the core of the jets is dominated by kinetic
energy flux ($Q < \, $1, blue region), while the bulk of the beams
is PFD ($Q > \, $1, red region).  Such distribution is consistent
with the one in the laboratory jets of \citet[][section~\ref{exps}]{lab1}.
The time average mean $Q$ of our magnetic tower beams is $\sim\,$6,
in agreement with the magnetic towers of \citep[][see their Fig.~3b,
bottom]{kato}. \\

\newpage

A very interesting result of our simulations is the long term
evolution of PFD jets which yields a series of collimated clumps,
the magnetization properties of which vary over time and distance
from the engine. PFD flows may thus eventually evolve into HD jets
at large distances from the central engine.  A future avenue for
this investigation is a comparison of the $Q({\bf X},t)$ distribution
between our magnetic towers --where $Q>$\,1 for the parameters
explored-- and models of jets created by MCL
processes; while MCL jets begin with $Q>\,$1 on scales less than
the Alfv\'en radius, in the asymptotic limit the kinetic energy
flux comes to dominate the flux of electromagnetic energy leading.
Simulations of MCL launching in which the flow is cold and gas
pressure can be ignored show typical values of $Q \sim \, $0.7 at
observationally-resolved distances from the engine (Krasnopolsky
et al. 1999, 2003). 

Changes in the distribution of $Q({\bf X},t)$ are quite relevant
for astrophysical jets. E.g. Hartigan et al. 2007 have shown that in YSO,
and perhaps in planetary nebulae jets as well, some mechanism may be
needed to reduce the magnetization of plasma close to the jet source.
If these flows began as magnetic towers then the disruption of the
central jets via kink modes may provide a means to produce collimated
high beta clumps of material as is observed in HH flows. Our
calculations are thus the fists steps towards distinguishing between
different launch mechanisms by providing descriptions of asymptotic
flow characteristics where observations might be possible, specially
with the next generation of telescopes, e.g. ALMA.

\section{Conclusions}

Our magnetic tower jet simulations are in good agreement with the
laboratory experiments of \citet{lab1}. In both investigations jets:
(1) carry axial
currents which return along the contact discontinuities, (2) the
cores have a high $\beta$ plasma, (3) the beams and cavities are PFD, 
(4) are eventually corrugated by
current driven instabilities becoming a collimated chain of 
magnetized ``clumps'' or ``bullets''.  We stress the
similarity between our models and the laboratory experiments because 
our implementation was not tuned to
represent the laboratory results at all.  
The laboratory
experiments and the simulations support each other then, as well as the
conclusion that both are revealing generic properties of PFD outflows.

Our simulations show that PFD jet beams are lighter, slower and
less stable than pre-collimated asymptotically hydrodynamic jets.
The latter might represent the asymptotic propagation regimes of
magneto-centrifugally launched jets. We find current-driven
perturbations in the PFD jets, which for the regimes studied are
amplified by cooling, firstly, and by base rotation, secondly. This
happens because shocks and thermal pressure support are weakened
by cooling, making the jets more susceptible to kinking. Base
rotation, on the other hand, amplifies the toroidal magnetic field
which in turn exacerbates the kink instability.  Eventually the
instabilities cause the corrugation of the jets' beam. PFD jets
become a collimated chain of magnetized ``clumps'' or ``bullets''
then.  \\

\textit{Financial support for this project was provided
by: the Space Telescope Science Institute grants HST-AR-11251.01-A
and HST-AR-12128.01-A; the National Science Foundation under
award AST-0807363 and grant PHY0903797; the Department of Energy 
under award DE-SC0001063; Cornell University grant 41843-7012;
the Laboratory for Laser Energetics of Rochester NY grant 
DE-FC52-08NA28302.}

\bibliographystyle{elsarticle-harv}



\end{document}